\documentclass[12pt,a4]{article}

\usepackage{amsthm,amsmath,latexsym,amssymb,amsfonts,amscd}
\usepackage{graphics,lscape,fancyhdr,array,stmaryrd,euscript}
\usepackage{empheq}
\usepackage{dsfont}
\usepackage{verbatim}
\usepackage{color,tikz,tikz-cd}
\usetikzlibrary[snakes]
\usepackage{relsize,slashed}
\numberwithin{equation}{section}
\usepackage{hyperref}
\usepackage{setspace}

\usepackage[numbers,sort&compress]{natbib}
\usepackage[nottoc]{tocbibind}

\newcommand{\besubeqs}{\begin{subequations}}
	\newcommand{\esubeqs}{\end{subequations}}

\begin{document}
	\hfill
	\vspace{-1.5cm}
	\vskip 0.05\textheight
	\begin{center}
		{\Large\bfseries 
			Note on one parameter subgroups of SO(3,2) classification}

		\vspace{0.4cm}
		
		\vskip 0.03\textheight
		
		 Iva \textsc{Lovrekovi\'c} 
		
		\vskip 0.03\textheight
		
		\vspace{5pt}
		{\em
			 Institute for Theoretical Physics, TU Wien, \\
			 Wiedner Hauptstr. 8-10, 1040 Vienna, Austria\\
			
		}
	\end{center}
	
	\vskip 0.02\textheight
	
	\begin{abstract}
		
		We study one-parameter subgroups of SO(3,2) and construct corresponding solutions of three-dimensional conformal gravity. The classification extends the SO(2,2) classification used in the construction of the BTZ black hole and identifies two additional classes of geometries.
	\end{abstract}
	
	\tableofcontents
	
	\section{Introduction}

	Studying gravity in three dimensions has attracted significant interest because many conceptual features of higher-dimensional gravity are already present, while the technical analysis is considerably simpler. An important feature is that gravity can be formulated as a gauge theory using the Chern–Simons action based on an underlying gauge algebra. When the gauge algebra is $so(2,2)=sl(2,\mathbb{R})\times sl(2,\mathbb{R})$ we recover Einstein gravity \cite{Witten:1988hc}.
	
		In three dimensions, gravity has been extensively studied from many perspectives. One line of investigation concerns the geometry of spinning black holes with matter coupling and a negative cosmological constant. It was shown that identifying points of anti–de Sitter space under a discrete subgroup of $SO(2,2)$ leads to black hole solutions. The classification of elements of the $so(2,2)$ Lie algebra  \cite{Banados:1992gq} allows one to classify the corresponding solutions, including the massless spinless black hole and the spinning black hole with maximal angular momentum.

	In this work we study one-parameter subgroups of the conformal group $SO(3,2)$ in three dimensions. Starting from the Chern–Simons formulation of conformal gravity as a gauge theory of the conformal group, we construct a non-complete set of solutions associated with different classes of one-parameter subgroups.

	These solutions in conformal gravity are classified according to the Killing vectors generating the corresponding discrete identifications. Each Killing vector defines a matrix characterized by its eigenvalues and Casimir invariants. The group $SO(3,2)$ contains two additional generators compared to $SO(2,2)$, leading to two new classes of one-parameter subgroups and, consequently, to new types of solutions relative to those arising from 
	$SO(2,2)$. These additional classes allow for the construction of distinct geometries. Similarly to the Petrov classification, this classification can be applied to physical systems with $SO(3,2)$ symmetry that arise from conformal theories more general than three-dimensional conformal gravity (for example, conformal higher-spin theories in 3D).
	Below we briefly review the Chern–Simons formulation of conformal gravity and list the one-parameter subgroups of 
	$SO(3,2)$. We present one example for each of the two new classes and a generalized solution within the class of the BTZ black hole.

	
	
	\section{ Chern-Simons framework}
	\label{sec:newchs}
	The standard Chern–Simons action is
	\begin{align}
		S[A]=\int \mathrm{Tr}\left[A \wedge dA +\frac{2}{3} A\wedge A\wedge A\right].
		\label{cs}
	\end{align}
	Here 
	$A$ is a Lie-algebra–valued one-form, while the gauge parameter is a Lie-algebra–valued zero-form. When the algebra is 
	$so(3,2)$, the action describes conformal gravity (CG) in three dimensions \cite{Horne:1988jf}. The action is also equivalent to the CG action when 
	$A$ is expressed in terms of the Lorentz connection.
	
	The equations of motion are
	\begin{align}
		F = dA + A \wedge A = 0 .
		\label{eomcs}
	\end{align}
	After fixing the gauge parameters following \cite{Horne:1988jf}, the equations of motion reduce to
	\begin{align}
		D_k W_{ij} - D_j W_{ik} = 0 ,
	\end{align}
	where $W_{ij}=R_{ij}-\frac{1}{4}g_{ij}R$.
	A distinctive feature of three dimensions is that these equations imply that spacetime is conformally flat. The action is invariant under conformal transformations.
	
	The equations of motion admit anti–de Sitter space as a solution. Starting from anti–de Sitter space, one can construct new solutions by performing identifications with respect to a discrete subgroup of 
	$SO(3,2)$. In the following we study one-parameter subgroups of $SO(3,2)$. A thorough analysis for 
	$SO(2,2)$ was carried out in \cite{Banados:1992gq}.

	\section{One parameter subgroups of SO(3,2)}
	
	{\bf Overview of the classification.     }
	Different classes of metrics of conformal gravity in three dimensions can be organized according to the Killing vectors they admit, and the first and third Casimir invariants that define each class.
	The Killing vectors are obtained from $\omega^{ab}J_{ab}$,  for a matrix $\omega^{ab}$ and $J^{ab}$ conformal Killing vectors.  We can define Killing vectors of SO(3,2)  as $J_{ab}$ with 
	\begin{align}
		J_{ab}=x_b\frac{\partial}{\partial x_a}-x_a\frac{\partial}{\partial x_b} 
	\end{align}
	for $x^a=(u,v,x,y,z)$. 
	
	The most general Killing vector is defined by $\frac{1}{2}\omega^{ab}J_{ab}$, while the matrix $\omega^{ab}$ inherits the symmetry of $J_{ab}$ which makes it antisymmetric $\omega^{ab}=-\omega^{ba}$. 
	This classification we can see in the Table 
	(\ref{tab:my_label}).
	\begin{table}[ht!]
		\centering
		\begin{tabular}{|c|c|c|c|}
			\hline         Type& Killing vector & First  Casimir & Third Casimir \\ \hline
			$  I_a$&$b(J_{23}+J_{01})-a(J_{12}+J_{03})$& $4(b^2-a^2)$ & $4(a^3-3ab^2)$\\
			$I_b$ & $\lambda_1 J_{12}+\lambda_2 J_{03}$ & $-2(\lambda_2^2+\lambda_1^2)$ & $2(\lambda_2^3+\lambda_1^3) $\\
			$I_c$ & $b_2 J_{23}+b_1 J_{01}$ & $2(b_1^2+b_2^2)$ &0 \\
			$I_d$ & $b_1 J_{23}+\lambda J_{04}$ & $2(b_1^2-\lambda^2)$ & $ 2\lambda(b_1^2-\lambda^2)$ \\
			$ II_a$ & $-\lambda(J_{03}+ J_{12})+ J_{01}-J_{02}-J_{13}+J_{23}$& $ -4\lambda^2$ & $4\lambda^2(3+\lambda)$\\ 
			$II_b$ & $(b-1)J_{01}+(b+1)J_{32}+J_{02}-J_{13}$ & $4b^2$ & 0\\
			$III_{+}$ & $J_{23}-J_{13}$  & 0& 0 \\
			$III_{-}$ & $J_{02}-J_{01}$ &0 & 0\\
			V & $J_{04}+2J_{12}$ & 0& -2\\ \hline
		\end{tabular}
		\caption{ The table shows types of one-parameter subgroups of SO(3,2). They are identified by the Killing vector and their Casimir invariants.}
		\label{tab:my_label}
	\end{table}
	
	The parameters $a,b,b_1,b_2,\lambda_1$ and $\lambda_2$ are real numbers defined by eigenvalues of $\omega^{ab}$. They are used in defining the type of the subgroup. Parameters $\lambda_1$ and $\lambda_2$ represent real eigenvalues, while $a$, $b$, $b_1$, and $b_2$, are parameters coming from complex eigenvalues.  
	Derivation of $\omega^{ab}$ in terms of eigenvalues is shown in Appendix, in which one can also find the details of the classification. 
	
	In comparison with one parameter subgroups of the $SO(2,2)$, we have two additional types. These are type $I_d$ and type $V$. In the case of $SO(2,2)$, types that are interesting for defining a black hole are $I_b$, $II_a$ and $III_+$, for which the eigenvalues of $\omega_{ab}$ are all real. They describe a general black hole, an extreme black hole with non-zero mass, and a ground state with zero mass, respectively.
	The general black hole has $|J|< Ml $ for $J$ angular momenta, $M$ mass  of the black hole and $l$ AdS radius. Expressing the inner and outer horizon radii, $r_-$ and $r_+$, as functions of $J$ and $M$, one finds that beyond the limit $J = Ml$ the eigenvalues become complex conjugates.
	This implies that the $|J|>Ml$ is described by the metric of the type $I_a$. However, by keeping the $|J|<Ml$ and setting mass to be negative, leads to two imaginary eigenvalues for $r_-$ and $r_+$ which belong to type $I_c$, possibly describing negative mass solutions \cite{Regge:1974zd}. 
	The new type that appears in conformal gravity, type $I_d$, has one real and one imaginary eigenvalue. Since it accommodates only one real horizon, and it can have one purely imaginary eigenvalue, it reminds of the cosmological solutions and, Lobachevksy type of solutions \cite{Bertin:2012qw}. Solutions with one eigenvalue that can belong to more than one type, need to have determined Casimir which clarifies it. 
	This is because global Lobachevsky has one real, non-zero horizon. The existence of one imaginary horizon is reminiscent of the rotating Lobachevsky solution.

	From the other known solutions of conformal gravity, in type $I_a$ we can classify the metrics which are 3D analogs of the MKR (Mannheim-Kazanas-Riegert) solution \cite{Mannheim:1988dj} in 4D, the OTT (Oliva-Tempo-Troncoso) solutions \cite{Oliva:2009hz} $ds^2=\frac{dr^2}{ar^2+br+c}-(ar^2+br+c)dt^2+r^2d\varphi^2$
	when we have general choice of parameters. 

	
	
	{\bf Condition for absence of closed timelike curves.    }
	Every Killing vector generates a one-parameter subgroup of AdS isometries. For a given Killing vector $\xi$, this can be expressed as
	\begin{equation}
		P \rightarrow e^{t\xi}P
	\end{equation}
	For $t=2n\pi$, where n is an integer, this map defines an identification subgroup. 
	
	
	The space obtained by quotienting with respect to the identification subgroup, i.e., by identifying points along a given orbit, inherits a well-defined metric from AdS. The resulting quotient space also solves the field equations under consideration.
	As a consequence of the identification, curves lying on the same orbit that connect two points in AdS become closed in the quotient geometry. For the causal structure of the quotient to be well defined, such closed curves must be neither timelike nor null. A necessary condition for the absence of closed timelike curves (CTCs) is
	\begin{align}
		\xi\cdot \xi>0 
	\end{align}
	This condition is in general not enough to guarantee that we will not have closed CTCs, however in this case it is sufficient \cite{Banados:1992gq}.
	In certain regions, the Killing vectors used in the identifications and responsible for the black hole geometries become timelike or null. Such regions must be removed from AdS spacetime in order for the identifications to be admissible.
	The resulting spacetime, denoted AdS$'$, is geodesically incomplete, as some geodesics would otherwise cross from $\xi \cdot \xi >0$ to $\xi\cdot \xi <0$.
	The hypersurface $\xi\cdot\xi=0$ then appears as a singular boundary in the causal structure, since continuation beyond it would generate closed timelike curves.
	It is therefore treated as a true singularity in the quotient.
	
	\subsection{Three important types of one-parameter subgroups of SO(3,2)}

	\subsubsection{Type $I_b$}
	The most interesting type in the classification of $SO(2,2)$ one parameter subgroups is $I_b$, because it describes BTZ black hole. For $SO(3,2)$ this type generalizes BTZ in following way. 
	
	To map the SO(3,2) embedding space
	\begin{align}
		ds^2 = du^2 + dv^2 - dx^2 - dy^2 - dz^2
	\end{align}
	obeying to the hyperboloid constraint
	\begin{align}
		u^2 + v^2 - x^2 - y^2 = l^2,
	\end{align}
	into a three-dimensional space with scaling symmetry, one can use the coordinate transformation
	\begin{align}
		u &= l \cosh(\mu) \sin(\lambda), & v &= l \cosh(\mu) \cos(\lambda), & x &= l \sinh(\mu) \cos(\theta),\\
		y &= l \sinh(\mu) \sin(\theta), & z &= l. &
	\end{align}
	Treating $l$ as a variable rather than a constant, the differential $dl^2$ emerging from $(u,v,x,y)$ cancels with $dz^2 = dl^2$, which is not possible to happen in the SO(2,2) case. This yields
	\begin{align}
		ds^2 = l^2 \Big(-\cosh^2(\mu) d\lambda^2 + d\mu^2 + \sinh^2(\mu) d\theta^2 \Big),
	\end{align}
	with non-constant $l$. The evaluation of $\xi \cdot \xi$ proceeds similarly to the SO(2,2) situation.
	
	For the choice of coordinates similarly as in $SO(2,2)$ case
	\begin{align}
		u &= l \sqrt{A(r)} \sinh(r_2 \varphi - r_1 t), & v &= l \sqrt{B(r)} \cosh(r_2 t - r_1 \varphi),\\
		x &= l \sqrt{A(r)} \cosh(r_2 \varphi - r_1 t), & y &= l \sqrt{B(r)} \sinh(r_2 t - r_1 \varphi),\\
		z &= l \sqrt{B(r) - A(r)},
	\end{align}
	with
	\begin{align}
		A(r) &= \frac{r^2 - r_1^2}{r_2^2 - r_1^2}, & B(r) &= \frac{r^2 - r_2^2}{r_2^2 - r_1^2}.
	\end{align}
	This leads to a metric analogous to the BTZ black hole in the $SO(2,2)$ case, except that $l$ remains non-constant. The $dl^2$ contribution cancels due to the $z$-component term, resulting in
	\begin{align}
		ds^2 = l^2 \Bigg( \frac{r^2 dr^2}{(r^2 - r_1^2)(r^2 - r_2^2)} + (-r^2 + r_1^2 + r_2^2) dt^2 - 2 r_1 r_2 dt d\varphi + r^2 d\varphi^2 \Bigg),\label{eq313}
	\end{align}
	which provides a conformal version of the standard BTZ line element. The identification $\varphi \rightarrow \varphi + 2 k \pi$ along the $\varphi$ direction generates conformal generalization of BTZ black hole. 
	
	The Killing vector for this type is \begin{align}\xi=\lambda_{1}J_{12}+\lambda_2J_{03}=\lambda_1(x\partial_v+v\partial_x)+\lambda_2(y\partial_u+u\partial_y),\end{align} 
	while the norm and the allowed region are determined from 
	\begin{align}
		\xi\cdot\xi=\lambda_1^2(v^2-x^2)+\lambda_2^2(u^2-y^2)>0.
	\end{align}
	We can express $v^2-x^2$ using the embedding of $SO(3,2)$
	in the five dimensional flat spacetime through the equation $-v^2-u^2+x^2+y^2+z^2=-l^2.$ Then the allowed region can be expressed as 
	\begin{align}
		\infty>u^2-y^2>-\frac{\lambda_1^2(l^2- z^2)}{(\lambda_2^2-\lambda_1^2)}. \label{type1b}
	\end{align}
	For comparison lets remember the allowed regions in the BTZ case. There, the allowed regions are determined by  $\xi_{BTZ}\cdot\xi_{BTZ}>0$ is $\frac{-r_-^2l^2}{r_+^2-r_-^2}<u^2-x^2<\infty$. The regions have three different types bounded by null surfaces. The region 1. is outer region with $u^2-x^2>l^2$. There is no $y=0$ in that region. The region 2. $0<u^2-x^2<l^2$ is an intermediate region. The region 3.  is $-\frac{r_-^2l^2}{r_+^2-r_-^2}<u^2-x^2<0$, the inner region which exists when $r_-\neq0$. 
	In the case of $SO(3,2)$ the regions are determined by (\ref{type1b}). We notice that this is similar as in BTZ case, however the difference is there is $z$ in addition to $l$ on the right side. If we choose $z$ to be of the definite sign we consider only one branch. For $z=0$ we reduce to the $SO(2,2)$ case. For $z\neq0$ we have 
	\begin{enumerate}
		\item Outer region case 1, in which $u^2-y^2>l^2-z^2$ for $|z|<|l|$. For $|z|>|l|$ we have $u^2-y^2>-f(z)$  for $f(z)=l^2-z^2$, a case 2 for outer region. Then $u^2-y^2<f(z)$ is bounded by $f(z)$. 
		\item The intermediate region is obtained when $0<u^2-x^2<l^2-z^2$, for $|z|<|l|$ this is analogous to intermediate region of BTZ black hole. For the case 2 of intermediate region we have $0<u^2-x^2<-f(z)$ which is not possible. The allowed region is $f(z)<u^2-x^2<0$. 
		\item The inner region for $|z|<|l|$ is analogously for BTZ $l^2-z^2<u^2-y^2<0$. When $|z|>|l|$ we obtain $-f(z)<u^2-y^2<0$ where $u^2-y^2<f(z)<0$.
	\end{enumerate}
	We show the graph for these regions in the Figure 1. We can also see that for $z=0$ the graph reduces to the graph obtained for one parameter subgroups of the $SO(2,2)$ group. 
	\begin{figure}
		\centering
		\includegraphics[width=0.5\linewidth]{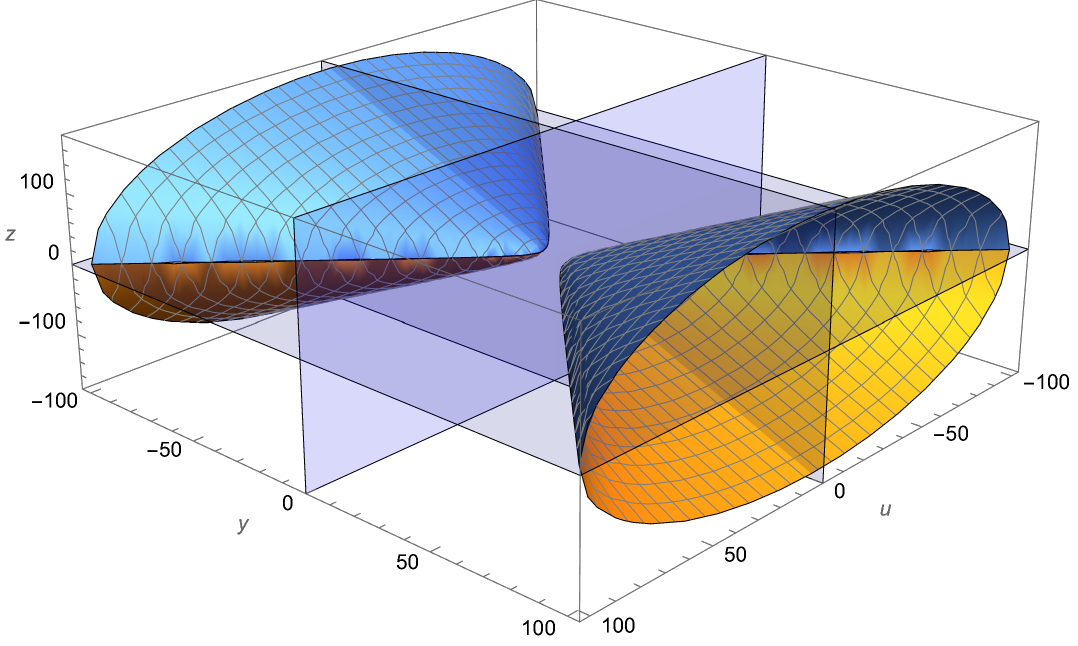}
		\caption{Regions obtained from the conformal Killing vector of type $I_b$. In  (\ref{type1b}) the $\lambda_1$ is taken to be 1, and $\lambda_2=2$, while $l$ is taken to be 10. The plotted surfaces denote $z$. $\lambda_1$ would correspond to $r_-$ in the BTZ picture when $z=0$ and $\lambda_2$ to $r_+$.}
		\label{fig1}
	\end{figure}
	
	\subsubsection{Type $I_d$}
	The new type of one-parameter subgroup in classification for $so(3,2)$ is type $I_d$.
	Here we consider the Killing vector for this type
	\begin{align}
		\xi=b_1 J_{23}+\lambda J_{04}= b_1(y\partial_x-x\partial_y)+\lambda(z\partial_u+u\partial_z) \label{xi1d}. 
	\end{align}
	The condition that there are no CTCS, $\xi\cdot \xi >0$, gives for the vector \newline $\xi_a=(\lambda z,0, b_1 y, -b_1 x,-\lambda u)$ in the  coordinates $(u,v,x,y,z)$ 
	the following conditions:
	\begin{align}
		\xi\cdot \xi=b_1^2(x^2-y^2)+\lambda^2(u^2-z^2),    
	\end{align}
	Using the hyperboloid constraint for $so(3,2)$ in which $z\neq l$ 
	$u^2+v^2-x^2-y^2-z^2=l^2$ we can write 
	\begin{align}
		\xi\cdot\xi =(u^2-z^2)(b_1^2+\lambda^2)+b_1^2v^2-l^2
	\end{align}
	The $\xi\cdot\xi>0$ region which is allowed can be written as 
	\begin{align}
		\infty>u^2-z^2>\frac{l^2}{b_1^2+\lambda^2}-\frac{b_1^2v^2}{b_1^2+\lambda^2} \label{ineq1d} 
	\end{align}
	It is instructive to examine this inequality for the specific regions
	and specific choices of the parameters $\lambda$ and $b_1$: 
	
	\begin{enumerate}
		\item We can write the first region similarly to exterior region in BTZ geometry. In our case we can have a region $u^2-z^2>l^2$ for $b_1^2+\lambda^2\sim1$ and very small $v$.  
		\item 
		Due to the fact that we kept $l^2$ and that we have conformal gravity case it looks as if we can write a region between the exterior and intermediate region 
		\begin{align}
			v^2>u^2-z^2>l^2
		\end{align}
		which is true for large $v$ and $b_1\text{ and }b_1^2+\lambda^2\sim1$. 
		\item An analog of intermediate region in BTZ geometry $l^2>u^2-z^2>0$, here appears for the $\frac{b_1^2v^2}{b_1^2+\lambda^2}\sim l^2$ and $b_1^2+\lambda^2>>l^2$.
		\item 
		The fourth region that we can consider is $0>u^2-z^2>\frac{l^2}{b_1^2+\lambda^2}-\frac{b_1^2v^2}{b_1^2+\lambda^2}$ which will hold when $\frac{l^2}{b_1^2+\lambda^2}<<1$. In this domain, the norm of the corresponding Killing vector is negative.
	\end{enumerate}
	
	\begin{figure}
		\centering
		\includegraphics[width=0.5\linewidth]{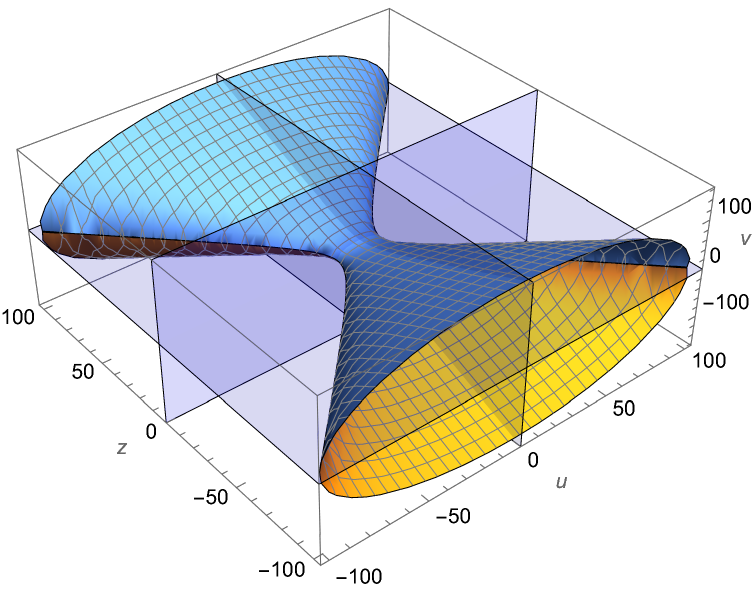}
		\caption{Regions from the conformal Killing vector of type $I_d$. In (\ref{ineq1d}) we use $\lambda=1.2$, $b_1=1$ and $l=20$. The plotted surfaces denote $v$.}
		\label{fig2}
	\end{figure}
	The vector characterizing the $I_d$ solutions can be expressed as $\xi_d = b_1 J_{23} + \lambda J_{04}$. Reducing from the 5D flat metric $-du^2 - dv^2 + dx^2 + dy^2 + dz^2$ with coordinates $(u,v,x,y,z)$, to 3D spacetime with coordinates $(r,t,\varphi)$, while preserving desired Killing vectors. \\ \\
	{\bf Case 1.} If we want to preserve the Killing vector $\partial_{\varphi}$ and $\partial_{t}$ we have to solve a set of partial differential equations to get conditions for compactifying the metric. However since we want $\partial_{\varphi}$ and $\partial_{t}$ conserved, we know the most general form of the metric in three dimensions which conserves this form. This is axially symmetric stationary metric.

	From the $\omega_{ij}$ matrix for $I_d$ in the appendix and the embedding space 
	\begin{align}
		ds^2 = du^2 + dv^2 - dx^2 - dy^2 - dz^2
	\end{align}
	of $SO(3,2)$, we can see that the eigenvalues are in two planes of the embedding space. Inspired by the ansatz in \cite{Banados:1992gq} for BTZ black hole, we can write for the first plane $(u,z)$
	\begin{align}
		u=r \sinh(t \lambda) && z=r\cosh(\lambda t)
	\end{align}
	Then we obtain $-du^2+dz^2=dr^2-r^2\lambda^2dt^2$.
	The remaining components we identify with 
	\begin{align}
		x=r \sin(b_1 t+\phi), && y=r \cos(b_1 t+\phi)
	\end{align}
	which leads to $-dr^2-r^2(b_1dt+d\phi)^2$. Taking $v=r$, from the embedding space we will obtain a form of the metric 
	\begin{align}
		ds^2=dr^2+(b_1^2-\lambda^2)r^2dt^2+2b_1r^2dtd\phi+r^2d\phi^2, \label{idsol}
	\end{align}
	where we had to multiply entire metric with -1. 
	The form (\ref{idsol}) that we obtained has Ricci scalar $R=-\frac{2}{r^2}$ and it satisfies Cotton tensor $\nabla_\sigma (R_{\mu\nu}-\frac{1}{4}g_{\mu\nu}R)=0$. If we multiply it with conformal factor $\frac{1}{r^2}$ we obtain flat metric. 
	In the picture of Einstein gravity (EG) it has the stress energy tensor $T_{rr}=\frac{1}{r^2}$. In the picture of ideal fluid this corresponds to pressureless dust which has diverging density in the center. It is a great example of a very simple form of the metric allowed in conformal gravity while not satisfying Einstein equations of motion. 
	\\ \\
	{\bf Case 2.} Preserving only the Killing vector $\partial_{\varphi}$, leads to a system of three partial differential equations   and $(u,v,x,y,z)$:
	\begin{align}
		\left(b_1(y\partial_x-x\partial_y)+\lambda(z\partial_u+u\partial_z) \right) r(u,v,x,y,z) &= 0,\\
		\left(b_1(y\partial_x-x\partial_y)+\lambda(z\partial_u+u\partial_z) \right) t(u,v,x,y,z) &= 0,\\
		\left(b_1(y\partial_x-x\partial_y)+\lambda(z\partial_u+u\partial_z) \right) \varphi(u,v,x,y,z) &= 1.
	\end{align}
	
	Solving this equations gives two solutions. It generally allows $r$ and $t$ to be expressed as
	\besubeqs
	\begin{align}
		r \to& f_r \Big[v,\frac{1}{2} \left(z^2-u^2\right), y \cos \left(\frac{b_1 \tanh ^{-1}\left(\frac{u}{z}\right)}{\lambda}\right)-x\left| \sin \left(\frac{b_1 \tanh ^{-1}\left(\frac{u}{z}\right)}{\lambda}\right)\right|, \nonumber \\& x \cos \left(\frac{b_1 \tanh ^{-1}\left(\frac{u}{z}\right)}{\lambda}\right)+y\left| \sin \left(\frac{b_1 \tanh ^{-1}\left(\frac{u}{z}\right)}{\lambda}\right) \right|\Big],\\
		t \to& f_t \Big[v,\frac{1}{2} \left(z^2-u^2\right), y \cos \left(\frac{b_1 \tanh ^{-1}\left(\frac{u}{z}\right)}{\lambda}\right)-x \left|\sin \left(\frac{b_1 \tanh ^{-1}\left(\frac{u}{z}\right)}{\lambda}\right)\right|, \nonumber \\& x \cos \left(\frac{b_1 \tanh ^{-1}\left(\frac{u}{z}\right)}{\lambda}\right)+y \left|\sin \left(\frac{b_1 \tanh ^{-1}\left(\frac{u}{z}\right)}{\lambda}\right)\right| \Big],
	\end{align}\label{idtrafo2}
	\esubeqs
	while the third equation fixes the conserved Killing vector along $\partial_\varphi$, giving
	\begin{footnotesize}
		\begin{align}
			\varphi \to& - \lambda \tanh ^{-1}\left(\frac{u}{|z|}\right)
			+ 2\lambda f_{\varphi}\Bigr[ v,\frac{1}{2} \left(z^2-u^2\right),y \cos \left(\frac{b_1 \tanh ^{-1}\left(\frac{u}{z}\right)}{\lambda}\right)-x\left| \sin \left(\frac{b_1 \tanh ^{-1}\left(\frac{u}{z}\right)}{\lambda}\right)\right|, \nonumber \\& x \cos \left(\frac{b_1 \tanh ^{-1}\left(\frac{u}{z}\right)}{\lambda}\right)+y \left|\sin \left(\frac{b_1 \tanh ^{-1}\left(\frac{u}{z}\right)}{\lambda}\right)\right|\Bigl]
			\label{fitrafo2}
		\end{align}
	\end{footnotesize}
	To insert these coordinates into the embedding metric $du^2+dv^2-dx^2-dy^2-dz^2$, one needs the inverse transformation $(u,v,x,y,z) \to (r,t,\varphi)$.
	In the case of BTZ black hole in the outer region, transformations similar to (\ref{idtrafo2}) and (\ref{fitrafo2}) take the form
	\besubeqs
	\begin{align}
		r &= \frac{1}{l}\sqrt{l^2 r_+^2 - (r_-^2 - r_+^2)(u-x)(u+x)},\\
		\varphi &= \frac{1}{r_-^2 - r_+^2}\left[r_+ \sinh^{-1}\left(\frac{v}{\sqrt{l^2 + u^2 - x^2}}\right) + r_- \cosh^{-1}\left(\frac{u}{\sqrt{u^2 - x^2}}\right)\right],\\
		t &= \frac{1}{r_-^2 - r_+^2}\left[r_- \sinh^{-1}\left(\frac{v}{\sqrt{l^2 + u^2 - x^2}}\right) + r_+ \cosh^{-1}\left(\frac{u}{\sqrt{u^2 - x^2}}\right)\right],
	\end{align}
	\esubeqs
	where $r_+$ and $r_-$ denote the outer and inner horizons, respectively. 
	
	The transformation (\ref{fitrafo2}) is selected in such a way that $\xi \propto \partial_\varphi$, which determines its form. 
	
	The straightforward example of the metric naturally comes from carefully selecting embedding coordinates and imposing the additional requirements. In the case of 3D conformal gravity this would be vanishing of the Cotton tensor.  That further restricts the allowed coordinate transformations.

	\subsubsection{Type $V$}
	
	Similarly as in the Case $I_d$ using the new basis using $\tilde{\omega}_{ij}$ we can construct a simpler form of the metric. Constructing a metric for $\omega_{ij}$ would be less transparent. 
	The Killing vector for this type is read out to be 
	\begin{align}
		\xi=J_{04}+2J_{12}=u\partial_z+z\partial_u+2(v\partial_x+x\partial_v)
	\end{align}. The scalar product of the Killing vectors 
	\begin{align}
		\xi\cdot\xi=u^2-z^2+4 (v^2- x^2)
	\end{align}
	gives the condition that the region $\xi\cdot\xi>0$ can be written as 
	\begin{align}
		\infty>(z^2-u^2)&>\frac{4}{3}(x^2-v^2)\\ 
		\infty>(z^2-u^2)&>-\frac{4}{3}(l^2+y^2).
	\end{align}
	Similarly like for the case $I_d$ we notice that in the inspection of regions, on the right hand side we obtain one extra component from the embedding. In the $I_d$ case the component on the right was $v$ component, while here it is $y$. If we think of the regular spacetime metric as one time component and three space components, (-,+,+,+), in this case for $so(3,2)$ we have (-,-,+,+,+) two time components and one three regular space components. In the $I_d$ case the conformal change of the regions can be thought as if it is done along the time component, while here, for case $V$, it happens along the space component. 
	The simplest transformation for the embedding coordinates we can recognize to be very similar to two cases
	\begin{itemize}\item To the case in $I_d$ case with two sets of hyperbolic functions defining embedding coordinates; and
		\item Since we have the structure of $\tilde{\omega}_{ij}$ similar to the structure of $\omega_{ij}$ in $I_b$ case (with $z$ and $y$ interchanged and fixed values for $\lambda_1$ and $\lambda_2$),  we are tempted to consider the similar construction. 
	\end{itemize}
	(i) We write the plane $(u,z)$ as in $I_d$ case
	\begin{align}
		u=r \sinh(t ) && z=r\cosh( t)
	\end{align}
	And the plane $v,x$ we define with 
	\begin{align}
		v=r \sinh(2t -\phi) && x=r\cosh(2 t-\phi).
	\end{align}
	Choosing this time $y=r$ the metric reads $ds^2=3 dr^2+4 r^2 dt d\phi -5 r^2 dt^2-r^2 d\phi ^2$. Ricci scalar of this solution is $-\frac{2}{3r^2}$. 
	\\
	(ii) The solution inspired by BTZ we read out from (\ref{eq313}) while choosing $y=l$
	\begin{align}
		ds^2 = l^2 \Bigg( \frac{r^2 dr^2}{(r^2 - 1)(r^2 -4)} + (-r^2 + 5) dt^2 - 4 dt d\varphi + r^2 d\varphi^2 \Bigg),\label{type5btz}
	\end{align}
	In the BTZ construction, the $I_b$ Killing vector is defined by $\xi_{BTZ}=\frac{r_+}{l}J_{12}+\frac{r_-}{l}J_{03}$. In case $V$  we can analogously think $\frac{r_+}{l}=2$ and $\frac{r_-}{l}=1$.
	However, this cannot be interpreted as BTZ with fixed horizons because the Casimir invariants do not match those of $I_b$ type. Here they are playing a decisive role that this type differes from $I_b$ and it is not a BTZ black hole with fixed horizons.

	\section{Conclusion}

	The three–dimensional solutions of conformal gravity have received considerably less attention than its Einstein–gravity counterpart. For this reason, we present a discussion of metrics associated with one–parameter subgroups of $SO(3,2)$. Our construction  followed the approach of \cite{Banados:1992gq}, where three–dimensional BTZ black holes were organized according to the one–parameter subgroups of $SO(2,2)$. In extending this framework to $SO(3,2)$, two additional types of solutions appeared that have no analogue in the $SO(2,2)$ classification.
	
	These new types, are cases $I_d$ and $V$, which required separate treatment. For both of them we explicitly constructed the coordinate transformations that allowed derivation of the corresponding line elements. In addition, we presented several explicit examples:
	(i) a representative geometry of type $I_d$ with one real and one imaginary eigenvalue,
	(ii) a type $I_b$ metric that can be interpreted as a conformal version of the BTZ black hole.
	(iii) And two type $V$ geometries inspired by the type $I_b$ and $I_d$ construction, however leading to a new type of metric.
	In Table 1, we summarized the Killing vectors for each type and presented their Casimir invariants. 
	
	The strategy of the classification is similar to the 
	Petrov classification \cite{Petrov:2000bs}, and similarly it can be used for theories that possess this symmetry.
	The first candidate is, of course, conformal gravity in 3D, whose solutions we consider here. Natural extensions are theories that include additional conformal fields and conformal higher spin theories \cite{Lovrekovic:2023xsj,Lovrekovic:2024yoo,Grigoriev:2019xmp,Pope:1989vj,Fuentealba:2024thk,Fuentealba:2020zkf}.
	In \cite{Lovrekovic:2023xsj} the theories with added higher spin fields have been considered with respect to which class of $SO(3,2)$ one-parameter subgroups they belong to.
	
	A complete geometric classification of the new types of solutions is left for future work.

	

	\addtocontents{toc}{\protect\setcounter{tocdepth}{1}}
	
	\section{Acknowledgements}
	The author thanks Daniel Grumiller and Evgeny Skvortsov for the  suggestions and discussions during the development and writing  of this work.
	This work was supported by the Hertha Firnberg grant T 1269-N and Elise Richter grant V 1052-N of the Austrian Science Fund FWF.
	
	\section{Appendix}
	

	
	Here we review how to obtain types of one-parameter subgroups. First we repeat the general construction, and then we construct each of the types in one-parameter subgroups of $SO(3,2)$. The construction for the types that appear in $SO(2,2)$ are very similar also here, while the constructions for the types $I_d$ and $V$ are new.  
	
	Two one-parameter subgroups $\{g(t)\}$ and $\{h(t)\}$, with $t\in \mathbb{R}$ are equivalent if and only if they are related by conjugation in G:
	$g(t)=k^{-1}h(t)k$ for $k\in G$.
	Equivalently, by rotating the  coordinates in $\mathbb{R}^4$ using the group G, we can map $g(t)$  on  $h(t)$. 
	Consequently, the classification problem amounts to organizing the elements of the Lie algebra 
	$g$ up to conjugacy.

	Elements of the Lie algebra $\mathfrak{g}$ are represented by real antisymmetric matrix $\omega_{ab} = - \omega_{ba}$. 
	Under conjugation of the infinitesimal transformation $R^a{}_b = \delta^a{}_b + \epsilon\, \omega^a{}_b$ by an element $k \in G$, the matrix $\omega_{ab}$ transforms according to
	\[
	\omega \;\longrightarrow\; \omega' = k^T \omega k, \qquad k \in G.
	\]
	\footnote{This is consistent with the condition $k^T \eta k = \eta$, where $\eta = \mathrm{diag}(- - + + +)$.}
	Which shows that we need to classify the antisymmetric matrices $\omega_{ab}$  up to this equivalence relation.

	We will use the Jordan--Chevalley decomposition, according to which any linear operator $M$ can be expressed as a sum
	\[
	M = S + N,
	\]
	where $S$ is semisimple linear operator \footnote{Semi-simple operator is defined as an operator diagonalizable over the complex numbers} and $N$ is nilpotent. Here,  $[S,N] = 0$ and $N^p = 0$ for some integer number $p$.
	The semisimple component can be written as $S = L^{-1} D L$ for an invertible matrix $L$ and a diagonal matrix $D$.
	This approach has an advantage that eigenvalues of $S$, are equal to eigenvalues of $M$, and characterize $S$.
	
	When degeneracies are not present, i.e. there is no repetitive eigenvalues, this implies $N = 0$, so that $M$ is classified up to similarity by its eigenvalues alone. 
	In the presence of degeneracies, however, the nilpotent contribution N becomes relevant. In order to reconstruct M we also need to know the details about nilpotent part, more precisely about  dimensions of the irreducible invariant subspaces.

	In what follows, we apply the Jordan--Chevalley decomposition to the operator $\omega^a{}_b$ and use it to organize the elements of the group $G$ under consideration. 
	The tensor $\omega^a{}_b$ is classified in a way analogous to the invariant classification of the electromagnetic field in Minkowski space.
	Because $\omega_{ab}$ is real and antisymmetric, the eigenvalues of its spectrum are restricted by the following constraints:
	\begin{itemize}
		\item if $\lambda$ is an eigenvalue of $\omega_{ab}$, then $-\lambda$ is also an eigenvalue;
		\item if $\lambda$ is an eigenvalue, then its complex conjugate $\lambda^*$ is as well an eigenvalue \cite{Banados:1992gq};
		\item since we are considering the group $SO(3,2)$, there is necessarily at least one vanishing eigenvalue, in comparison to what happens with $SO(2,2)$
	\end{itemize}

	Compared to the classification of eigenvalue types for $SO(2,2)$ given in \cite{Banados:1992gq}, the present case requires the inclusion of zero as an additional eigenvalue.
	As a result, the possible types take the following form:
	\begin{enumerate}
		\item $\lambda,-\lambda,\lambda^*,-\lambda^*$ and $0$, where $\lambda = a_1 + i a_2$ with $a_1 \neq 0$ and $a_2 \neq 0$,
		\item $\lambda_1 = \lambda_1^*, -\lambda_1, \lambda_2 = \lambda_2^*, -\lambda_2$, and $0$, with $\lambda_1$ and $\lambda_2$ real,
		\item $\lambda_1, \lambda_1^* = -\lambda_1, \lambda_2, \lambda_2^* = -\lambda_2$, and $0$, where $\lambda_1$ and $\lambda_2$ are purely imaginary,
		\item $\lambda_1 = \lambda_1^*, -\lambda_1, \lambda_2, \lambda_2^* = -\lambda_2$, and $0$, with $\lambda_1$ real and $\lambda_2$ purely imaginary.
	\end{enumerate}
	Each type contains two independent real parameters together with one vanishing eigenvalue.
	
	Degenerate cases appear in the following situations:
	\begin{itemize}
		\item For $\lambda \neq 0$ the roots are generically distinct, whereas for $\lambda = 0$ all five roots are the same.
		\item In cases (2) and (3), if $\lambda_1 = \lambda_2$ or $\lambda_2 = -\lambda_1$, degeneracies appear even when $\lambda_1 \neq 0$; if $\lambda_1 = 0$ this again leads to a quintopole root.
		\item In cases (2), (3), and (4), whenever one of the eigenvalues vanishes, there appears triple zero root.
	\end{itemize}

	Assuming that the classification principles for $SO(2,2)$ extend to $SO(3,2)$, one may use $SO(3,2)$ transformations to bring $\omega_{ab}$ into a canonical form uniquely determined by its set of eigenvalues whenever the eigenvalues are simple.
	In this situation, $\omega_{ab}$ can be written in a basis in which $\omega^a{}_b$ is diagonal.
	When eigenvalues are degenerate, however, distinct canonical forms may appear due to the presence of a nontrivial nilpotent contribution $N$ contained in $\omega^a{}_b$.
	In such cases, one must determine a canonical form for each allowed structure of $N$.
	
	Below we list these canonical forms.
	Following this convention, we refer to $\omega^{ab}$ as being of type $k$ if its nilpotent component satisfies $N^k = 0$.

	We continue to classify the operators according to their Jordan--Chevalley structure.\\
	(Type I, corresponding to $N = 0$, coincides with the Type I classification for $SO(2,2)$ with an additional zero eigenvalue):
	\begin{enumerate}
		\item $I_a$: four complex eigenvalues plus zero
		\item $I_b$: four real eigenvalues plus zero
		\item $I_c$: four purely imaginary eigenvalues plus zero
		\item $I_d$: two real and two imaginary eigenvalues, together with zero
	\end{enumerate}
	
	Type II corresponds to $N \neq 0$ with $N^2 = 0$:
	\begin{enumerate}
		\item $II_a$: zero, plus two real double roots $\lambda$ and $-\lambda$
		\item $II_b$: zero, plus two imaginary double roots
		\item $II_c$: one triple root at zero and two simple roots $\lambda$ and $-\lambda$, where $\lambda$ may be real or imaginary
	\end{enumerate}
	
	Type III has $N^2 \neq 0$, $N^3 = 0$, with a single quintuple zero root.
	
	Type IV satisfies $N^3 \neq 0$, $N^4 = 0$, also with a quintuple zero root.
	
	Type V has $N^4 \neq 0$, $N^5 = 0$, again yielding a quintuple zero root.
	
	We will require the following property of the eigenvectors of $\omega^i{}_j$: if $v^i$ and $u^i$ correspond to eigenvalues $\lambda$ and $\mu$, then
	\[
	v_a u^a = 0 \quad \text{unless } \lambda + \mu = 0.
	\]
	Moreover, whenever $\lambda \neq 0$, the eigenvector $v^a$ is null.

	\subsection{Type $I_a$}
	
	Following the definition of Type $I_a$, the operator $\omega_{ij}$ satisfies
	\begin{align}
		\omega_{ab} l^b &= \lambda l_a,\\
		\omega_{ab} m^b &= -\lambda m_a,\\
		\omega_{ab} l^{*b} &= \lambda^* l_a^*,\\
		\omega_{ab} m^{*b} &= -\lambda^* m_a^*,\\
		\omega_{ab} k^b &= 0.
	\end{align}
	
	To examine whether additional nonzero scalar products exist beyond $l^a m_a = l^{*a} m^*_{a} = 1$, we consider $k^a l_a$. A term of the form $k_a l_b + l_a k_b$ in the metric would correctly lead to $k_b$ after contraction with $k^a$, but contracting with $m^a$ would give $l_a = l_a + k_a$, forcing $k_a = 0$. Therefore, the only non-vanishing contribution from $k_a$ can arise via the scalar product $k_a k^a$. The resulting metric then reads
	\begin{align}
		\eta_{ij} = l_{(i} m_{j)} + l^*_{(i} m^*_{j)} + k_{(i} k_{j)}.
	\end{align}
	
	Expressing the complex vectors in terms of their real and imaginary parts, $l_a = u_a + i v_a$ and $m_a = n_a + i q_a$, the metric becomes
	\begin{align}
		\eta_{ij} = 2 \left( u_{(i} n_{j)} - v_{(i} q_{j)} \right) + k_{(} k_{j)}.
	\end{align}
	
	The spin connection retains the same structure as in the $so(2,2)$ case, with $\lambda = a + i b$:
	\begin{align}
		\omega_{ij} = 2 a \left( u_{[i} n_{j]} - v_{[i} q_{j]} \right) - 2 b \left( u_{[i} q_{j]} + v_{[i} n_{j]} \right).
	\end{align}
	
	In an orthonormal basis with components
	\[
	u_a = \left(0, \tfrac12, \tfrac12, 0,0\right), \quad
	n_a = \left(0, -\tfrac12, \tfrac12, 0,0\right), \quad
	v_a = \left(\tfrac12, 0, 0, \tfrac12 ,0\right), \quad
	q_a = \left(\tfrac12, 0, 0, -\tfrac12 ,0\right),
	\]
	the $\omega_{ij}$ takes the form
	\begin{align}
		\omega_{ij} =
		\begin{pmatrix}
			0 & b & 0 & a & 0 \\
			-b & 0 & a & 0 & 0 \\
			0 & -a & 0 & b & 0 \\
			-a & 0 & -b & 0 & 0 \\
			0 & 0 & 0 & 0 & 0
		\end{pmatrix}.
	\end{align}
	
	The Casimir invariants are given by
	\begin{align}
		I_1 &= 4(-a^2 + b^2),\\
		I_3 &= 4(a^3 - 3 a b^2),
	\end{align}
	where $I_1$ corresponds to those in the $so(2,2)$ case.

	\subsection{Type $I_b$}
	
	The extension of Type $I_b$ from $so(2,2)$ to $so(3,2)$ proceeds similarly for the nonzero eigenvalues. By definition,
	\begin{subequations}
		\begin{align}
			\omega_{ij} l^j &= \lambda_1 l_i, & \omega_{ij} m^j &= -\lambda_1 m_i,\\
			\omega_{ij} n^j &= \lambda_2 n_i, & \omega_{ij} u^j &= -\lambda_2 u_i,\\
			\omega_{ij} k^j &= 0.
		\end{align}
	\end{subequations}
	The main difference arises in the metric, which acquires an additional term $k_a k_b$. The matrix $\omega_{ab}$ in an orthonormal basis keeps the $so(2,2)$ form with an extra row and column to accommodate $k_a$. Explicitly, the metric and $\omega_{ab}$ are
	\begin{align}
		\eta_{ij} &= l_i m_j + m_i l_j + n_i u_j + u_i n_j + k_i k_j,\\
		\omega_{ij} &= \lambda_1 (l_i m_j - m_i l_j) + \lambda_2 (n_i u_j - u_i n_j).
	\end{align}
	\begin{align}
		\omega_{ab} =
		\begin{pmatrix}
			0 & 0 & 0 & -\lambda_2 & 0 \\
			0 & 0 & -\lambda_1 & 0 & 0 \\
			0 & \lambda_1 & 0 & 0 & 0 \\
			\lambda_2 & 0 & 0 & 0 & 0 \\
			0 & 0 & 0 & 0 & 0
		\end{pmatrix}.
	\end{align}
	
	The associated Casimir invariants take the form
	\begin{align}
		I_1 &= -2 \left( \lambda_1^2 + \lambda_2^2 \right),\\
		I_3 &= 2 \left( \lambda_1^3 + \lambda_2^3 \right).
	\end{align}
	Notice that in the Type $I_a$ and Type $I_b$ cases exactly one eigenvalue vanishes, whereas the remaining eigenvalues are nonzero and coincide with those appearing in the $SO(2,2)$ classification.

	\subsection{Type $I_c$}
	
	For Type $I_c$, the set of eigenvalues of $\omega_{ab}$ consists entirely of purely imaginary eigenvalues:
	\begin{align}
		\omega_{ij} l^j &= i b_1\, l_i, & \omega_{ij} l^{j*} &= - i b_1\, l_i^*, \\
		\omega_{ij} m^j &= i b_2\, m_i, & \omega_{ij} m^{j*} &= - i b_2\, m_i^*, \\
		\omega_{ij} k^j &= 0. &
	\end{align}
	The only nonzero scalar contractions are $l_i l^{i*}$, $m_i m^{i*}$, and $k_i k^i$.
	These define the metric
	\begin{align}
		\eta_{ij} = l_i l_j^* + l_j l_a^* + m_i m_j^* + m_j m_i^* + k_i k_j.
	\end{align}
	
	The corresponding $\omega_{ij}$ is given by
	\begin{align}
		\omega_{ij} = i b_1 ( l_i l_j^* - l_j l_i^* ) - i b_2 ( m_i m_j^* - m_j m_i^* ).
	\end{align}
	
	Introducing real vectors via $l^a = \tfrac{1}{\sqrt{2}}(u^a + i v^a)$, one finds that in an orthonormal basis $\omega_{ij}$ can be written as
	\begin{align}
		\omega_{ij} =
		\begin{pmatrix}
			0 & b_1 & 0 & 0 & 0 \\
			- b_1 & 0 & 0 & 0 & 0 \\
			0 & 0 & 0 & b_2 & 0 \\
			0 & 0 & - b_2 & 0 & 0 \\
			0 & 0 & 0 & 0 & 0
		\end{pmatrix}.
	\end{align}
	
	The  Casimir invariants in this case are
	\begin{align}
		I_1 &= 2(b_1^2 + b_2^2), \\
		I_3 &= 0.
	\end{align}

	\subsection{Type $I_d$}
	
	The Type $I_d$  represents the first non-trivial extension of the $\mathrm{SO}(2,2)$ classification to the $\mathrm{SO}(3,2)$ case. We write the eigenvalue relations
	\begin{align}
		\omega_{ij}l^j&=\lambda l_i,  &&  \omega_{ij}n^j=-\lambda n_i \\
		\omega_{ij}m^j&=ib_1m_i, && \omega_{ij}m^{j*}=-ib_1m_i^* .
	\end{align}
	The scalar products 
	\[
	l\cdot l=n\cdot n=m\cdot m=m^*\cdot m^*=0,
	\]
	vanish, while the only non-vanishing scalar products are
	\[
	l\cdot n=1, \qquad m\cdot m^*=\pm1.
	\]
	In the $\mathrm{SO}(2,2)$ case this restricts the admissible signatures of $\eta_{ij}$. The allowed forms are $(+-++)$ and $(+---)$, 
	excluding the $(--++)$ signature. 
	
	The enlargement to $\mathrm{SO}(3,2)$ introduces an additional zero eigenvalue
	\begin{align}
		\omega_{ij}k^j=0,
	\end{align}
	with $k\cdot k\neq0$. The metric may then be written as
	\begin{align}
		\eta_{ij}=l_in_j+l_jn_i+m_im_j^*+m_jm_i^*+k_ik_j.
	\end{align}
	Introducing the decomposition
	\[
	m_i=\frac{1}{\sqrt{2}}(u_i+iv_i),
	\]
	where the scalar product of the real vectors is  $u\cdot u=v\cdot v=\pm1$ and $u\cdot v=0$.
	
	In this decomposition and using the basis vectors 
	$l=\frac{1}{\sqrt{2}}(1,0,0,0,1), n=\frac{1}{\sqrt{2}}(-1,0,0,0,1),u={0,0,0,1,0},v={0,0,-1,0,0}$ and $k=(0,1,0,0,0)$ we can write the metric as
	\begin{align}
		\eta_{ij}=l_in_j+l_jn_i+u_iu_j+v_jv_i-k_ik_j. \label{metric1d}.
	\end{align}
	The $\omega_{ij}$ is more convenient to be represented in this basis and it takes the form
	\begin{align}
		\omega_{ij}&=\lambda(l_in_j-n_il_j)+ib_1(m_im_j^*-m_jm_i^*) \\ 
		&=\lambda(l_in_j-n_il_j)+b_1(u_jv_i-u_iv_j),
	\end{align}
	which is in our chosen orthonormal basis represented by
	\begin{align}
		\omega_{ij}=\left(\begin{array}{ccccc}
			0 & 0 & 0 & 0 & \lambda \\
			0 & 0 & 0 & 0 & 0 \\
			0 & 0 & 0 & b_1 & 0 \\
			0 & 0 & -b_1 & 0 & 0 \\
			-\lambda & 0 & 0 & 0 & 0
		\end{array}\right). \label{omega1d}
	\end{align}
	The corresponding Casimir invariants are
	\begin{align}
		I_1&=2(b_1^2-\lambda^2),\\
		I_3&=2\lambda(b_1^2-\lambda^2).
	\end{align}
	
	
	\subsection{Type $II_a$}
	
	This class is characterized by two equal non-vanishing double eigenvalues together with a single zero eigenvalue. The presence of the additional null eigenvalue does not change $\omega_{ab}$ beyond the changes already encountered in the Type $I$ cases. 
	
	It can be written 
	\begin{align}
		\omega_{ij}l^j&=\lambda l_i, && \omega_{ij}u^j=\lambda u_i+l_i,\\
		\omega_{ij}m^j&=-\lambda m_i, && \omega_{ij}s^j=-\lambda s_i+m_i,\\
		\omega_{ij}k^j&=0. &&
	\end{align}
	From these relations one infers the metric and the $\omega_{ij}$ in the form
	\begin{align}
		\eta_{ij}=l_{(i}s_{j)}-m_{(i}u_{j)}+k_{(i}k_{j)},\\
		\omega_{ij}=\lambda\bigl(l_{[i}s_{j]}-m_{[i}u_{j]}\bigr)-l_{[i}m_{j]}.
	\end{align}
	In a basis adapted to this decomposition, the $\omega_{ij}$ is represented by
	\begin{align}
		\omega_{ij}=\left(\begin{array}{ccccc}
			0 & 1 & 1 & \lambda & 0 \\
			-1 & 0 & \lambda & 1 & 0 \\
			-1 & -\lambda & 0 & 1 & 0 \\
			-\lambda & -1 & -1 & 0 & 0 \\
			0 & 0 & 0 & 0 & 0
		\end{array}\right).
	\end{align}
	The corresponding Casimir invariant  are
	\begin{align}
		I_1&=-4\lambda^2,\\
		I_3&=4\lambda^2(3+\lambda).
	\end{align}

	\subsection{Type $II_b$}
	
	This class corresponds two purely imaginary nonzero double eigenvalues together with a single vanishing eigenvalue. It may be viewed as the straightforward analogue of the $so(2,2)$ construction, supplemented by the additional zero root.
	
	Here, we can write
	\begin{align}
		\omega_{ij}l^j&=ib\,l_i, && \omega_{ij}u^j=ib\,u_i+l_i,\\
		\omega_{ij}l^{j*}&=-ib\,l_i^*, && \omega_{ij}u^{j*}=-ib\,u_i^*+l_i^*,\\
		\omega_{ij}k^j&=0. &&
	\end{align}
	The corresponding non-vanishing inner products give for the metric
	\begin{align}
		\eta_{ij}=-l_i^*u_j-l_j^*u_i+l_iu_j^*+l_ju_i^*+k_ik_j.
	\end{align}
	The associated $\omega_{ij}$ may be expressed as
	\begin{align}
		\omega_{ij}=ib\bigl(l_i^*u_j-l_j^*u_i+l_iu_j^*-l_ju_i^*\bigr)+l_i^*l_j-l_j^*l_i.
	\end{align}
	In a basis $\omega_{ij}$ becomes
	\begin{align}
		\omega_{ij}=\left(\begin{array}{ccccc}
			0 & b-1 & -1 & 0 & 0 \\
			-b+1 & 0 & 0 & 1 & 0 \\
			1 & 0 & 0 & b+1 & 0 \\
			0 & -1 & -b-1 & 0 & 0 \\
			0 & 0 & 0 & 0 & 0
		\end{array}\right).
	\end{align}
	The Casimir invariants for this class evaluate to
	\begin{align}
		I_1&=4b^2,\\
		I_3&=0.
	\end{align}

	\subsection{Type $II_c$}
	
	By knowing our eigenvalues, here we can write
	\begin{align}
		\omega_{ij}l^j&=0, && \omega_{ij}m^j=l_i\\
		\omega_{ij}u^j&=\lambda u_i, && \omega_{ij}v^j=-\lambda v_i \\\omega_{ij}k^j&=0 &&
	\end{align}
	
	In this case we obtain the metric which is degenerate. The reason for this is that one of the eigenvectors would need to have vanishing scalar product $v\cdot v=0$, while being non-zero and orthogonal to all the other vectors. This kind of construction would give degenerate metric.

	\subsection{Type $III_a$}
	
	In this type the characteristic equation has an quintuplet eigenvalue. As in the earlier cases, the $so(3,2)$ realization closely parallels the corresponding $so(2,2)$ construction, with the extra zero root included.
	
	The equations for $\omega^i{}_j$ are
	\begin{align}
		\omega_{ij}l^j&=0,\\
		\omega_{ij}m^j&=0, \qquad \omega_{ij}u^j=m_i, \qquad \omega_{ij}t^j=u_i,\\
		\omega_{ij}k^j&=0.
	\end{align}
	The inner products satisfy $m\cdot m= m\cdot u=0$, while $m\cdot t\neq0$, together with $l\cdot l\neq0$ and $k\cdot k\neq0$. We fix these by writing
	\[
	l\cdot l=\epsilon_1=\pm1,\qquad k\cdot k=\epsilon_2 .
	\]
	With this normalization the metric takes the form
	\begin{align}
		\eta_{ij}=\epsilon_1\bigl(-l_il_j-m_it_j-t_jm_i+u_iu_j\bigr)+\epsilon_2 k_i k_j.
	\end{align}
	The corresponding $\omega_{ab}$ reduces to
	\begin{align}
		\omega_{ij}=\epsilon_1\,(m_i u_j-u_i m_j).
	\end{align}
	Since the sign $\epsilon_1$ distinguishes inequivalent realizations, one obtains two cases. For $\epsilon_1=+1$ one has $III_{a+}$,
	\begin{align}
		\omega_{ij}=\left(
		\begin{array}{ccccc}
			0 & 0 & 0 & 0 & 0 \\
			0 & 0 & 0 & 1 & 0\\
			0 & 0 & 0 & 1 & 0\\
			0 & -1 & -1 & 0 & 0\\
			0 & 0 & 0 & 0 & 0
		\end{array}
		\right),
	\end{align}
	while for $\epsilon_1=-1$ one obtains $III_{a-}$,
	\begin{align}
		\omega_{ij}=\left(
		\begin{array}{ccccc}
			0 & -1 & -1 & 0 & 0\\
			1 & 0 & 0 & 0 & 0 \\
			1 & 0 & 0 & 0 & 0 \\
			0 & 0 & 0 & 0 & 0\\
			0 & 0 & 0 & 0 & 0 
		\end{array}
		\right).
	\end{align}
	In both cases the Casimir invariants vanish.

	\subsection{Type $III_{b}$}
	
	Here one obtains
	\begin{align}
		\omega_{ij}l^j&=0, && \omega_{ij}k^j=l_i, && \\
		\omega_{ij}m^j&=0, && \omega_{ij}u^j=m_i && \omega_{ij}t^j=u_i.
	\end{align}
	For the scalar products we have
	$l\cdot l= l \cdot k= l \cdot m=l\cdot u=m \cdot m=m\cdot u=m \cdot k=u\cdot t = t \cdot u=0$. The non-vanishing scalar products are $l \cdot t=-u \cdot k$ and $m\cdot t=-u\cdot u$. 
	Before looking at additional contractions, consider the scalar product $m\cdot t$. One could perform a redefinition of the form $m^i \rightarrow m^i + l^i$, which allows this inner product to be set to zero. This redefinition, however, also makes $m$ orthogonal to all basis vectors, including itself, so that $m\cdot m=0$ as well. As a consequence, the resulting metric would become degenerate, and such a choice must therefore be excluded.
	
	\subsection{Type $III_c$}
	Here, we can write
	\begin{align}
		\omega_{ij}l^j&=0, && \omega_{ij}m^j=l_i, && \omega_{ij}k^j=m_i\\
		\omega_{ij}u^j&=\lambda u_i, && \omega_{ij}v^j=-\lambda v_i &&.
	\end{align}
	In this case we obtain a degenerate metric because the 
	scalar product of $l$ with the rest of the vectors and itself vanishes.

	\subsection{Type $IV$}
	
	Analogously as for the $so(2,2)$ group, here, type $IV$ is also inconsistent.

	\subsection{Type $V$}
	
	For this type, the equation of $\omega_{ij}$ acting on the basis vectors can be expressed as
	\begin{align}
		\omega_{ij}l^j&=0, && \omega_{ij}m^j=l_i, && \omega_{ij}u^j=m_i, && \omega_{ij}t^j=u_i, & \omega_{ij}k^j=t_i.
	\end{align}
	The  scalar products that vanish are
	\[
	l\cdot l, \; l\cdot m, \; l\cdot u, \; l\cdot t, \; m\cdot m, \; u\cdot m, \; m\cdot k, \; u\cdot t, \; k \cdot t.
	\]
	We may fix
	$
	l\cdot k=\epsilon, m\cdot t=-u\cdot u=-\epsilon,
	$
	and notice that $u\cdot k=-t\cdot t$, which requires a redefinition to achieve $u\cdot k=0$. Introducing
	$u^i \rightarrow u^i + l^i,$ we have shifted $u^i$ and fixed $u\cdot k=0$ which automatically fixes $t\cdot t=0$. The metric becomes
	\begin{align}
		\eta_{ij}=u_i u_j+m_i t_j+t_i m_j+l_i k_j+k_i l_j.
	\end{align}
	The matrix $\omega_{ab}$ consistent with the eigenvalues is 
	\begin{align}
		\omega_{ij}=(l_i t_j-t_i l_j)+(m_i u_j-u_i m_j),
	\end{align}
	which explicitly  reads
	\begin{align}
		\omega_{ij}=\left(
		\begin{array}{ccccc}
			0 & \frac{1}{2} & 0 & \frac{1}{2} & 0 \\
			-\frac{1}{2} & 0 & \frac{1}{2} & 0 & -1 \\
			0 & -\frac{1}{2} & 0 & -\frac{1}{2} & 0 \\
			-\frac{1}{2} & 0 & \frac{1}{2} & 0 & 1 \\
			0 & 1 & 0 & -1 & 0 \\
		\end{array}
		\right),
	\end{align}
	for the choice of vectors
	\[
	u=(0,0,0,0,1),\; m=(0,-1,0,1,0),\; k=(\frac{1}{2},0,\frac{1}{2},0,0),\; l=(-1,0,1,0,0),\; t=(0,-\frac{1}{2},0,-\frac{1}{2},0).
	\]
	If we build a matrix $M$ from the basis vectors $(l,m,u,k,t)$, we can transform the matrix to dual basis $\tilde{\omega}=M \omega M^T$ to obtain a simpler form 
	\begin{align}
		\tilde{\omega}_{ij}=\left(
		\begin{array}{ccccc}
			0 & 0 & 0 & 0 & 1 \\
			0 & 0 & 2 & 0 & 0 \\
			0 & -2 & 0 & 0 & 0 \\
			0 & 0 & 0 & 0 & 0 \\
			-1 & 0 & 0 & 0 & 0 \\
		\end{array}
		\right).
	\end{align}
	The Casimir invariants are
	\begin{align}
		I_1&=0,\\
		I_3&=-2.
	\end{align}
	
	


\end{document}